\documentstyle[preprint,aps,epsfig]{revtex}

	{\arrayclosep 0.14em\begin{eqnarray}}{\end{eqnarray}}
\newfont{\Fr}{eufm10}   
\def\be{\begin{equation}}
\def\ee{\end{equation}}
\def\bea{\begin{eqnarray}}
\def\eea{\end{eqnarray}}
\def\b{\begin{eqnarray*}}
\def\e{\end{eqnarray*}}

\def \( {\left(}
\def \) {\right)}

\def\[{\left[}
\def\]{\right]}

\def\Journal#1#2#3#4{{#1}{\bf#2}, #4 (#3)}

\def\MEETtmp#1#2#3#4{{#1} {\it #2}, {#3}, {(#4)}}
\def\NPB{{ Nucl. Phys.} \bf B}
\def\NPBPS{{ Nucl. Phys.} {\bf B} (Proc. Suppl.) }
\def\PLB{{ Phys. Lett.}  {\bf B} }

\def\PRD{{ Phys. Rev.} {\bf D} }
\def\PRC{{ Phys. Rep.} \bf C}

\def\PTPS{{ Prog. Theor. Phys. Suppl. }}

\begin{document}
\draft
\title{Dual Higgs Mechanism based on the Dual Gauge Formalism 
 in the Lattice QCD}
\author{Atsunori Tanaka and Hideo Suganuma}
\address{Research Center for Nuclear Physics (RCNP), Osaka University\\
Mihogaoka 10-1, Ibaraki, Osaka 567-0047, Japan \\
E-mail: suganuma@rcnp.osaka-u.ac.jp}
\maketitle
\begin{abstract}
We
study the dual Higgs mechanism induced by monopole condensation
based on
the dual gauge formalism
in the maximally abelian (MA) gauge
in the lattice QCD.
To examine ``monopole condensation''
in QCD, we study the monopole part or
the monopole-current system appearing in the MA gauge
by extracting the dual gluon field $B_\mu$.
First, 
we investigate the inter-monopole potential using
the dual Wilson loop
in the lattice QCD simulation.
In the monopole part in the MA gauge,
the inter-monopole potential is found to be 
flat, and can be fitted as
the Yukawa potential in the infrared region.
From more detailed analysis of
the inter-monopole potential
considering the monopole size,
we estimate the effective dual-gluon mass $m_B \simeq 0.5$GeV 
and the effective monopole size $R_{^{_{\rm M}}} \simeq 0.2$fm.
Second, we study the dual gluon propagator
$G^D_{\mu\nu}(x-y) \equiv \langle B_\mu(x) B_\nu(y) \rangle_{\rm MA}$ 
in the MA gauge,
and find that
$G^D_{\mu\mu}$
behaves as the massive vector-boson
propagator with $m_B \simeq 0.4$ GeV
in the infrared region.
The effective-mass acquirement of the dual gluon field $B_\mu$
at the long distance
can be regarded as the lattice QCD evidence of
``infrared monopole condensation'' in the MA gauge.
\end{abstract}
\vspace{0.5cm}

\vfill\eject
\section{Introduction}
Quantum chromo-dynamics (QCD) is
the fundamental theory of the strong interaction
and exhibits the interesting phenomena
such as color confinement and
dynamical chiral-symmetry breaking (D$\chi$SB)
\cite{Pokorski,Cheng,Itzykson,Huang,Aitchison,Greiner,Rothe}.
The QCD Lagrangian has the SU($N_c$) local symmetry
and is described by the quark field $q$ and the gluon field $A_\mu$ as
\begin{equation}
{\cal L}_{QCD}=-\frac{1}{2}{\rm tr}(G_{\mu\nu}G^{\mu\nu})
+\bar{q}(i/\!\!\!\!D - m_q)q,
\end{equation}
where $G_{\mu\nu}$ denotes the SU($N_c$) field strength
$G_{\mu\nu} \equiv \frac{1}{ie}[{D}_\mu, {D}_\nu]$
with the covariant-derivative
operator ${D}_\mu \equiv {\partial}_\mu
+ ieA_\mu$ \cite{Pokorski}.
In QCD, 
the asymptotic freedom is one of the most important features in QCD
\cite{Pokorski,Cheng,Itzykson},
and the QCD gauge-coupling constant $e$
becomes small in the ultraviolet region.
On the other hand, in the low-energy region,
the gauge-coupling constant $e$ becomes large,
and there arise the nonperturbative-QCD (NP-QCD)
phenomena like color confinement
and D$\chi$SB
corresponding to the strong-coupling nature.
These NP-QCD phenomena
are extremely difficult to understand
in the analytical manner
from QCD and
have been studied by using the effective models
\cite{Greiner}
or the lattice QCD simulation \cite{Rothe}.

In the lattice formalism,
space-time coordinates are discretized with the lattice spacing $a$,
and the theory is described by the link variable 
$U_\mu(s)\equiv e^{iaeA_\mu(s)} \in$ SU($N_c$).
For instance, the standard lattice action is given as
\begin{eqnarray}
S=\beta \sum_{s, \mu>\nu}[1-\frac{1}{2N_c}{\rm tr}
\{U_{\mu\nu}(s)+U^\dagger_{\mu\nu}(s) \}],
~\beta \equiv \frac{2N_c}{e^2}
\end{eqnarray}
where $U_{\mu\nu}(s)$ is the plaquette variable defined as
$U_{\mu\nu}(s) 
\equiv U_\mu(s)U_\nu(s+\mu)U^\dagger_\mu(s+\nu)U^\dagger_\nu(s)$.
The lattice QCD Monte Carlo simulation is 
based on the numerical
calculation of the QCD partition functional,
and it is one of the most reliable methods directly
from QCD.
In the lattice QCD,
nonperturbative quantities
like the quark confinement potential,
the chiral condensate
$\langle {\bar q}q\rangle$
and low-lying hadron masses are well reproduced \cite{Rothe}.
However, one cannot understand the confinement mechanism
only by looking on the result obtained by the lattice QCD.

Recently, the lattice QCD simulation
has shed light on the confinement mechanism
in terms of the dual-superconductor
picture,
which was first proposed by
Nambu,'t~Hooft and Mandelstam
in the middle of 1970's
\cite{Nambu,tHooft75,Mandelstam}.
In this scenario,
quark confinement can be understood
with the dual version of the superconductivity.
In the ordinary superconductor, the Meissner effect
occurs by condensation
of the Cooper-pair with the electric charge
\cite{Abrikosov}.
Consider existence of the magnetic charges with the
opposite sign immersed in the superconductor,
then the magnetic flux is squeezed like a tube
between the magnetic charges, and
the magnetic potential between them
becomes linear as the result of the Meissner effect
\cite{Abrikosov}.
On the other hand,
the confinement force between the color-electric charge
is characterized by the universal physical quantity
of the string tension $\sigma = 0.89 \sim 1$ GeV/fm,
which is evaluated
from the Regge trajectory of hadrons and the lattice QCD simulations.
This string tension
is brought by one-dimensional squeezing of the color-electric
flux in the QCD vacuum \cite{Polikarpov}.
In the dual-superconductor scenario \cite{Nambu},
the QCD vacuum is assumed as the dual version of the superconductor,
and the dual Meissner effect
brings the one-dimensional flux tube \cite{Haymaker}
between the quark and the anti-quark,
which leads to the linear confinement potential 
\cite{Huang,Aitchison,Greiner,Rothe,Nambu,SST}.

The dual Higgs mechanism, however, requires
``color-magnetic monopole condensation'' as
the dual version of electric condensation
in the superconductor,
although QCD dose not include
{\rm the color-magnetic monopole}
as the elementary degrees of freedom.
On the appearance of magnetic monopoles from QCD,
't~Hooft showed that
QCD is reduced to an abelian gauge theory with magnetic monopoles
by taking the abelian gauge,
which fixes the partial gauge symmetry SU($N_C$)/U$(1)^{N_c-1}$
through the diagonalization of
a gauge-dependent variable \cite{tHooft81}.
Here,
the color-magnetic monopole appears as the topological object
corresponding to the nontrivial homotopy group
$\pi_{2}$(SU($ N_{c}$)/U$(1)^{N_{c}-1}$)=$Z_{\infty}^{N_{c}-1}$.

Recent lattice QCD studies show abelian dominance \cite{Ezawa},
which means that NP-QCD phenomena
like confinement \cite{Yotsuyanagi,Hioki91}
and
D$\chi$BS \cite{Miyamura,Woloshyn}
are almost described only by the diagonal gluon component
 in the maximally abelian (MA) gauge
\cite{Kronfeld,Brandstater}.
The MA gauge is a sort of the abelian gauge where off-diagonal
gluon components
are minimized by the gauge transformation.
In the MA gauge,
physical information of the gauge configuration is maximally concentrated
into the diagonal gluon component.
As for the NP-QCD phenomena,
QCD can be well approximated by the abelian projected QCD
(AP-QCD),
which is the abelian gauge theory
extracted from QCD in
the MA gauge by removing the off-diagonal gluon \cite{Kronfeld,Brandstater}.

AP-QCD includes not only electric
currents $j_\mu$ but also magnetic currents $k_\mu$,
and can be decomposed into the photon part and the monopole part
corresponding to the separation of $j_\mu$ and $k_\mu$,
respectively \cite{DeGrand,Stack_Wensley}.
In other words, AP-QCD consists of the two parts.
The photon part of AP-QCD (photon-projected QCD) is
the abelian gauge theory only with electric currents $j_\mu$
like the ordinary QED.
On the other hand,
the monopole part of AP-QCD (monopole-projected QCD)
only includes monopole currents $k_\mu$.
The lattice QCD studies show monopole dominance,
which means that NP-QCD phenomena are almost
described only by the monopole projected QCD
\cite{Miyamura,Stack_Wensley,Bali,Polikarpov,STSM,Sasaki_Miyamura}.
In particular, the lattice QCD simulation numerically shows that
the monopole current $k_\mu$ is responsible for the electric confinement
and $j_\mu$ does not contribute to it
in the MA gauge \cite{Stack_Wensley,Bali,Polikarpov}.
Since we are interested in the infrared physics such as confinement and
its mechanism in the QCD vacuum,
we discard the off-diagonal gluon and the photon part in the MA gauge
as the irrelevant ingredients,
because they do not contribute to the quark confinement.
Then, we concentrate ourselves on the monopole part in the MA gauge
for the study of the confinement mechanism.


Then, the remaining problem is
whether ``monopole condensation''
occurs in the QCD vacuum or not.
In this paper,
we perform the SU(2) lattice QCD simulation in the MA gauge,
and extract the monopole current $k_\mu$
as the relevant degrees of freedom for confinement.
Then, we numerically derive the dual gluon field $B_\mu$
from $k_\mu$ 
\cite{INNOCOM_Tanaka,YKIS_Suganuma}, and
investigate
the effective-mass acquirement of the dual gluon $B_\mu$
in the QCD vacuum by examining
the inter-monopole potential and the dual gluon propagator.

\newpage
\section{Abelian Projection and Monopole Projection
in the lattice QCD formalism}

\subsection{Abelian-Projected QCD in the MA gauge}

Recent studies with
the lattice QCD Monte Carlo simulation
have revealed the abelian dominance and the monopole dominance
in the maximally abelian (MA) gauge
for the nonperturbative QCD (NP-QCD) phenomena
such as confinement, dynamical chiral-symmetry breaking
and instantons \cite{Yotsuyanagi,Hioki91,Miyamura,Woloshyn,DeGrand,Stack_Wensley,Bali,Polikarpov,STSM,Sasaki_Miyamura}.
In the SU(2) lattice formalism,
the MA gauge fixing is achieved by maximizing
\begin{equation}
R=
\frac{1}{2}{\rm tr}\sum_{s, \mu}
[\tau^{3}U_\mu (s){\tau^3}U^{\dagger}_\mu (s)]
=\sum_{s, \mu}\left[1-2 \left(
\{ U_\mu^{1}(s) \} ^{2}+ \{ U_\mu^{2}(s) \} ^{2}
\right) \right]
\label{eq:MAfix}
\end{equation}
by the SU(2) gauge transformation,
\begin{equation}
U_\mu \to U^{\rm MA}_\mu=V(s)U_\mu(s)V^\dagger(s+{\mu}), 
\label{eq:MA-tr}
\end{equation}
where $V(s)$ and $V(s+{\mu})$
are the gauge functions located at the starting 
and end points of the link variable $U_\mu(s)$.
In this gauge, 
the absolute value of off-diagonal components $U^1_\mu(s)$
and $U^1_\mu(s)$ 
are forced to be small as possible
using the gauge degrees of freedom.

In accordance with the Cartan decomposition, 
the SU(2) link variable $U_\mu(s)$ is factorized as 
\begin{eqnarray}
U^{\rm MA}_{\mu}(s)&=&M_{\mu}(s)u_{\mu}(s)~,\\
M_{\mu}&\equiv&
\exp[i(\theta_{\mu}^{1}\tau^{1}+\theta_{\mu}^{2}\tau^{2})],
~~u_{\mu}(s)\equiv \exp[i(\theta_{\mu}^{3}\tau^{3})],
\label{eq:M-def}
\end{eqnarray}
where $u_{\mu} \in {\rm U(1)_{3}}$
and $M_{\mu} \in {\rm SU(2)/U(1)_{3}}$ 
correspond to 
the diagonal part and the off-diagonal part of the gluon field,
respectively.
In the continuum limit,
the angle variable $\theta^a_\mu$ goes to the gluon field $A^a_\mu$
as 
$\theta^a_\mu \to \frac{1}{2}eaA^a_\mu$.
The off-diagonal factor $M_\mu(s)$
is rewritten as
\begin{eqnarray}
M_{\mu}(s)=
e^{i(\theta_{\mu}^{1}\tau^{1}+\theta_{\mu}^{2}\tau^{2})}
&=&
{\left(
\begin{array}{cc}
\cos \theta_{\mu}^{\rm } & -\sin \theta_{\mu}^{\rm }e^{-i\chi_{\mu}} \\
\sin \theta_{\mu}^{\rm }e^{i\chi_{\mu}} & \cos \theta_{\mu}^{\rm }
\end{array}
\right)}.
\end{eqnarray}
with
\begin{equation}
\theta_{\mu}^{\rm } \equiv {\rm mod}_{\pi/2}
\sqrt{(\theta_{\mu}^{1})^{2}+(\theta_{\mu}^{2})^{2}},
~~~~~\chi_\mu \equiv \tan^{-1}\frac{\theta_{\mu}^{1}}{\theta_{\mu}^{2}}.
\nonumber
\end{equation}
Here, the parameter ranges are usually taken as
\begin{equation}
\theta_\mu \in [0, \frac{\pi}{2}],
~\chi_\mu \in [0, 2\pi),
~\theta^3_\mu \in [0, 2\pi),
\end{equation}
which provides the one-to-one correspondence
to the SU(2) group element.
Under the abelian gauge transformation with $v(s) \in {\rm U(1)_{3}}$,
$M_\mu(s)$ and $u_{\mu}(s)$ are transformed as
\begin{eqnarray}
M_{\mu}(s) \to M^v_{\mu}(s)=v(s)M_{\mu}(s)v^{\dagger}(s),\\
u_{\mu}(s) \to u^v_{\mu}(s)=v(s)u_{\mu}(s)v^{\dagger}(s+{\mu}),
\end{eqnarray}
to  keep the form of Eq.(\ref{eq:M-def})
for $M^v_\mu \in$ SU(2)/U$(1)_3$ and $u^v_\mu \in$ U$(1)_3$.
Then, $M_\mu(s)$ behaves as the charged matter field
and the abelian link-variable
\begin{equation} 
u_{\mu}(s)=
\left(
\begin{array}{cc}
e^{i\theta_{\mu}^{3}} & 0 \\
0 & e^{-i\theta_{\mu}^{3}} 
\end{array}
\right)
\end{equation}
behaves as a abelian gauge field
with respect to the residual abelian gauge symmetry.


As a remarkable feature of the MA gauge,
the abelian dominance holds for the NP-QCD phenomena
such as quark confinement and chiral-symmetry breaking
\cite{Yotsuyanagi,Hioki91,Miyamura,Woloshyn}.
Here, we call abelian dominance for an operator $\hat O$,
when the expectation value $\langle O[U_\mu] \rangle$
is almost equal to the expectation value 
$\langle O[u_\mu] \rangle_{\rm MA}$, 
where off-diagonal gluons are dropped off 
in the MA gauge.
For instance, the abelian string tension 
$\sigma_{\rm Abel}\equiv\langle \sigma(U_\mu) \rangle_{\rm MA}$
in the MA gauge is almost equal to 
$\sigma_{\rm SU(2)}\equiv\langle \sigma(U_\mu) \rangle$
as $\sigma_{\rm Abel} \simeq 0.92\sigma_{\rm SU(2)}$ for $\beta \simeq 2.5$
in the lattice QCD 
\cite{Bali,Polikarpov}.
Thus, NP-QCD phenomena 
are almost reproduced only by the abelian link variable $u_{\mu}$,
 and off-diagonal gluon components $\theta^1_\mu$, $\theta^2_\mu$
 do not contribute to NP-QCD in the MA gauge.
Hence, as long as the infrared physics is concerned,
QCD in the MA gauge
can be approximated by the abelian projected QCD (AP-QCD),
where the SU(2) gluon field $\theta^a_\mu \tau^a$ is replaced by
the abelian gluon field $\theta^{3}_\mu \tau^3$.
In other words, AP-QCD is the abelian gauge theory
keeping essence of NP-QCD, and is extracted from
QCD in the MA gauge.
Hereafter, we pay attention to the AP-QCD described
by $\theta^{3}_\mu$ in the MA gauge.

\subsection{Decomposition of AP-QCD into Monopole Part and Photon Part}
The abelian-projected QCD (AP-QCD)
includes not only the color-electric current $j_\mu$ 
but also the color-magnetic monopole current $k_\mu$ \cite{tHooft81}.
In the Maxwell equations with $j_\mu$ and $k_\mu$,
the field strength $\theta_{\mu\nu}^{\rm Abel}$ satisfies as 
\begin{eqnarray}
\partial_{\mu}\theta^{\rm Abel}_{\mu\nu}&=&j_\nu\\
\partial_{\mu}{~^{*}\!\theta^{\rm Abel}_{\mu\nu}}
&=&k_\nu
\end{eqnarray}
with $^*\!\theta^{\rm Abel}_{\mu\nu}
\equiv \frac12 \varepsilon_{\mu\nu\rho\sigma}
\theta^{\rm Abel}_{\rho\sigma}$.
In the presence of both $j_\mu$ and $k_\mu$,
the field strength $\theta^{\rm Abel}_{\mu\nu}$ 
cannot be described by the simple two-form
$\partial_\mu \theta_\nu^{\rm 3} - \partial_\nu \theta_\mu^{3}$
 with the regular one-form $\theta^{3}_\mu$ \cite{Blagojevic}.
Using the diagonal gluon component $\theta^3_\mu$, 
the abelian field strength $\theta^{\rm Abel}_{\mu\nu}$ is defined by
\begin{eqnarray}
\theta_{\mu\nu}^{\rm Abel} =
{\rm mod}_{2\pi}(\partial_\mu\theta_\nu^{\rm 3}
-\partial_\nu\theta_\mu^{\rm 3}) ~~\in [-\pi, \pi), 
\end{eqnarray}
which is U$(1)_3$-gauge invariant.
In the continuum limit $a \to 0$,
$\theta^{\rm Abel}_{\mu\nu}$ goes to the abelian field strength $F_{\mu\nu}$
in the physical unit as 
$\theta^{\rm Abel}_{\mu\nu} \to \frac{1}{2}ea^2F_{\mu\nu}$,
and therefore the range of $\theta^{\rm Abel}_{\mu\nu}$
is required to include $0$.
Then, the two-form 
$\partial_\mu \theta_\nu^{\rm 3} - \partial_\nu \theta_\mu^{3}$
can be decomposed as 
\begin{equation}
\partial_\mu\theta^{\rm 3}_\nu
-\partial_\nu\theta^{\rm 3}_\mu
= \theta^{\rm Abel}_{\mu\nu}+2\pi n_{\mu\nu},
\label{eq:field2}
\end{equation}
where the former part denotes the field strength
and the latter part $2\pi n_{\mu\nu}(s) \in 2\pi {\bf Z}$
corresponds to the Dirac string on the lattice \cite{DeGrand}.
Thus, in the ordinary description 
\cite{Blagojevic},
the system includes the 
singularity as the Dirac string $2 \pi n_{\mu\nu}$,
which makes the analysis complicated.


To clarify the roles of $j_\mu$ and $k_\mu$
to the nonperturbative quantities of QCD,
we consider the decomposition of AP-QCD
into the photon part
and the monopole part,
corresponding to the separation of $j_\mu$ and $k_\mu$.
We call this separation into the photon and monopole parts
as the ``photon projection'' and the ``monopole projection'', respectively
\cite{STSM,HI_HS}.

In the lattice formalism, the photon part
$\theta^{\rm Ph}_\mu(s)$ and the monopole part
$\theta^{\rm Mo}_\mu(s)$ are
obtained from $\theta^{\rm Abel}_{\mu\nu}(s)$ 
and $2\pi n_{\mu\nu}(s)$, respectively,
\begin{eqnarray}
\theta^{\rm Ph}_{\mu}(s)&=&
\{\Box^{-1}\partial_{\nu}\theta^{\rm Abel}_{\mu\nu}\}(s) 
\label{eq:L-def-ph}\\
\theta^{\rm Mo}_{\mu}(s)&=&
2\pi \{\Box^{-1}\partial_{\nu}n_{\mu\nu}\}(s),
\label{eq:L-def-mo}
\end{eqnarray}
using 
the inverse d'Alembertian $\Box^{-1}$ on the lattice 
\cite{Miyamura,QULEN_Ichie}.
Here,
$\Box^{-1}$ is the non-local operator \cite{DeGrand,QULEN_Ichie},
\begin{eqnarray}
\langle s|\Box^{-1}|s' \rangle
= -\frac{1}{4 \pi^2}\frac{1}{(s-s')^2}
\end{eqnarray}
which satisfies
\begin{equation}
\Box_s \langle s|\Box^{-1}| s' \rangle=\delta^4(s-s')=\langle s|s' \rangle.
\end{equation}
The diagonal gluon component $\theta^{\rm 3}_\mu(s)$ is 
found to be decomposed as 
\begin{equation}
\theta^{\rm 3}_\mu(s)=\theta^{\rm Ph}_\mu(s)+\theta^{\rm Mo}_{\mu}(s)
\label{eq:decom2}
\end{equation}
in the Landau gauge, $\partial_{\mu}\theta^{\rm 3}_{\mu}(s)=0$. 
The field strengths, $\theta^{\rm Ph}_{\mu\nu}$ in the photon part
and $\theta^{\rm Mo}_{\mu\nu}$ in the monopole part,
are given as
\begin{eqnarray}
\theta^{\rm Ph}_{\mu\nu} =
{\rm mod_{2\pi}}(\partial_\mu\theta^{\rm Ph}_\nu-\partial_\nu
\theta^{\rm Ph}_\mu) ~~\in [-\pi, \pi) \\
\theta^{\rm Mo}_{\mu\nu} =
{\rm mod_{2\pi}}(\partial_\mu\theta^{\rm Mo}_\nu-\partial_\nu
\theta^{\rm Mo}_\mu) ~~\in [-\pi, \pi)
\label{eq:field-m}
\end{eqnarray}
on the lattice, respectively.

In the actual lattice QCD simulation,
the monopole current $k_\mu$ and 
the electric current $j_\mu$ are slightly modified through
the monopole  and the photon projections, respectively,
due to the numerical error on the lattice.
However, these differences are negligibly small in the actual
lattice QCD simulation.
In fact, $k_\mu^{\rm Mo} \simeq k_\mu$ and $j_\mu^{\rm Mo} \simeq 0$
hold in the monopole part, and $j_\mu^{\rm Ph} \simeq j_\mu$
and $k_\mu^{\rm Ph} \simeq 0$ hold in the photon part within 1$\%$ error
\cite{STSM}. 
Here, we have kept the labels as ``$Mo$'' and ``$Ph$'' 
for the electric current and the monopole current,
and we have used $(k_{\mu}^{\rm Mo},j_{\mu}^{\rm Mo})$
and $(k_{\mu}^{\rm Ph},j_{\mu}^{\rm Ph})$ for these currents
in the monopole part and the photon part, respectively.

As a remarkable fact,
lattice QCD simulations show that nonperturbative 
quantities such as the string tension,
the chiral condensate and instantons 
 are almost reproduced only by the monopole part
in the MA gauge, which is called as monopole dominance 
\cite{Stack_Wensley,Bali,Polikarpov,STSM}. 
On the other hand,
the photon part dose not contribute 
these nonperturbative quantities in QCD.

{\it 
Since we are interested in the NP-QCD phenomena,
it is convenient and transparent to
extract the relevant degrees of freedom for NP-QCD
by removing irrelevant degrees of freedom like 
the off-diagonal gluons ($\theta^1_\mu$, $\theta^2_\mu$)
and the electric current $j_\mu$.
Therefore, we
concentrate ourselves to the monopole part,
which keeps the essence of NP-QCD such as confinement,
and we examine whether monopole condensation
occurs or not.
In other words, our aim is to investigate the feature
of the monopole-current system appearing in the MA gauge.}

\newpage
\section{Dual Gauge Formalism}
In the MA gauge,
the monopole part carries essence of the
nonperturbative QCD like the electric confinement
\cite{Miyamura,Stack_Wensley,Bali,Polikarpov,STSM,Sasaki_Miyamura}.
Since the monopole part only
includes the color-magnetic current $k_\mu$, i.e. $j_\mu$=0,
the Maxwell equation in the monopole part
becomes
\begin{eqnarray}
\partial_{\mu}\theta^{\rm Mo}_{\mu\nu}&=&0\\
\partial_{\mu}{^{*}\!\theta^{\rm Mo}_{\mu\nu}}&=&k_{\nu},
\end{eqnarray}
where $\theta^{\rm Mo}_{\mu\nu}$ denotes 
the field strength in the monopole part.
This system resembles the dual version of QED
with $j_\mu \ne 0$ and $k_\mu=0$, and hence
it is useful to introduce
the dual gluon field $B_{\mu}(s)$ instead of $\theta^{\rm Mo}_\mu(s)$ 
in the monopole part
for the study of the dual Higgs mechanism in QCD
\cite{INNOCOM_Tanaka}.

The dual gluon field $B_{\mu}$ is defined so as to satisfy
the definition of the abelian gauge field,
\begin{equation}
\partial_\mu B_\nu - \partial_\nu B_\mu=^*\!\theta^{\rm Mo}_{\mu\nu},
\label{eq:def-bmu}
\end{equation}
which is the dual version of the ordinary relation,
$F_{\mu\nu} \equiv \partial_\mu A_\nu - \partial_\nu A_\mu$
\cite{YKIS_Suganuma}.
The interchange between $A_\mu$ and $B_\mu$
corresponds to the electro-magnetic duality transformation,
${\bf H} \leftrightarrow {\bf E}$.
Therefore, owing to the absence of $j_\mu$,
the dual gauge field $B_\mu$
can be introduced without the singularity
like the Dirac string.
In other words,
the absence of $j_\nu$ is automatically derived as
the dual Bianchi identity,
\begin{equation}
j_\nu=\partial_\mu \theta_{\mu\nu}
=\partial_{\mu}~^*\!(\partial \land B)_{\mu\nu}=0.
\label{eq:bianchi}
\end{equation}

Let us consider the derivation of the
dual gauge field $B_\mu$ from the monopole current $k_\mu$.
Taking the dual Landau gauge $\partial_{\mu}B_\mu=0$,
the relation
$\partial_\mu~^*\!\theta^{\rm Mo}_{\mu\nu}
=\partial^2 B_\nu - \partial_\nu(\partial_\mu B_\mu)
=k_\nu$ becomes a simple form $\Box B_\mu=k_\mu$.
Therefore,
the dual gluon field $B_\mu(x)$
is obtained
as
\begin{eqnarray}
B_{\nu}(x)=\left( \Box^{-1}k_{\nu} \right)(x)
&=& \int d^4y \langle x |\Box^{-1}|y \rangle k_\nu (y).
=-\frac{1}{4\pi^2} \int d^4y \frac{k_\nu(y)}{(x-y)^2}
\end{eqnarray}
by using the inverse d'Alembertian $\Box^{-1}$ \cite{DeGrand,QULEN_Ichie}.
Here, the dual gauge formalism provides the natural description of
 the monopole part in terms of the monopole current $k_\mu$
 and the dual gluon $B_\mu$.
Moreover, the dual gauge formalism is useful
to examine the dual Higgs mechanism in QCD
\cite{INNOCOM_Tanaka,YKIS_Suganuma}.

In the dual superconductor picture in QCD,
$k_\mu$ and $B_\mu$ correspond to
the Cooper-pair and the photon 
in the superconductor, respectively.
The Cooper-pair and the photon are essential
degrees of freedom which bring the superconductivity.
In the superconductor,
the photon field $A_\mu$
gets the effective mass
as the result of Cooper-pair condensation,
and this leads to the Meissner effect. 
Accordingly, the potential between the static electric charges
becomes the Yukawa potential $V_Y(r) \propto \frac{e^{-mr}}{r}$
in the ideal superconductor obeying the London equation.
Similarly,
the dual gluon $B_\mu$ is expected to be massive
in the  the monopole-condensed system,
and the mass acquirement of $B_\mu$
leads to the dual Meissner effect.
In other words,
the acquirement of dual gluon mass $m_B$
reflects monopole condensation,
and brings electric confinement.
Hence, we can investigate the dual Higgs mechanism in QCD
by evaluating the dual gluon mass $m_B$,
which is evaluated from the inter-monopole potential.

To estimate the inter-monopole potential,
we propose the dual Wilson loop $W_D$ 
\cite{INNOCOM_Tanaka,YKIS_Suganuma}.
The dual Wilson loop $W_D$ is defined 
by the line-integral of the dual gluon field 
$B_\mu$
along a closed loop $C$,
\begin{equation}
W_D(C) \equiv
Re[\exp{(i \oint_C B_\mu dx_\mu)}],
\label{eq:def-dw}
\end{equation}
which is the {\it dual version of the abelian Wilson loop} 
$
W_{\rm Abel}(C)
\equiv
Re[\exp{(i\oint_C \theta_\mu^{3}dx_\mu)}].
$
Here, the dual Wilson loop
$W_D(R \times T)$
describes the interaction between
the monopole-pair with the test magnetic charges
$\frac{e}{2}$ and $-\frac{e}{2}$.
The inter-monopole potential\footnote{
Rigorously, $V_M(r)$ corresponds to the inter-monopole potential
in the monopole part
(the monopole-current system)
in the MA gauge.}
is obtained from the dual Wilson loop as
\begin{equation}
V_{M}(R) = -\lim_{T \to \infty} \frac{1}{aT}\ln 
\langle W_D(R \times T)\rangle
\label{eq:imp}
\end{equation}
in a similar manner to the extraction of the inter-quark potential
from the Wilson loop 
\cite{Huang,Aitchison,Greiner,Rothe}.

To summarize, starting from the monopole current $k_\mu$ in the MA gauge,
we have introduced the dual gluon field $B_\mu$
and the dual Wilson loop
for the investigation of the dual Higgs mechanism in QCD.
\newpage
\section{Inter-Monopole Potential and Dual Gluon Effective Mass}
In this section, we show the numerical result of 
the lattice QCD simulation.
For the study of the dual Higgs mechanism in QCD,
we calculate the dual Wilson loop $W_D(R,T)$ and 
investigate the inter-monopole potential $V_M(r)$ in the MA gauge
using the SU(2) lattice with $20^4$ and $\beta=2.2 \sim 2.3$.
All measurements are performed at every 100 sweeps
after a thermalization of 5000 sweeps using 
the heat-bath algorithm.
We prepare 100 samples of gauge configurations.
These simulations have been performed using NEC SX-4 at 
Computation Center of Osaka University.


In the dual Higgs mechanism, it is essential that 
the dual gluon field $B_\mu$ acquires the effective mass.
For investigating the effective-mass acquirement of the dual gluon
$B_\mu$ which is brought by the monopole condensation,  
let us extract the inter-monopole potential $V_M(r)$
from the dual Wilson loop $\langle W_{D}(R,T) \rangle_{\rm MA}$
obtained by the lattice QCD.
The dual Wilson loop $\langle W_{D}(R,T) \rangle_{\rm MA}$
seems to obey the perimeter law rather than 
the area law
for large loops as shown 
in Fig.\ref{fig:dw}.  
Since the dual Wilson loop $\langle W_{D}(R,T) \rangle_{\rm MA}$
 satisfies the perimeter law as
\begin{equation}
\ln \langle W_{D}(R \times T) \rangle_{\rm MA} \simeq -2a(R+T)\cdot \alpha
\label{eq:per-dw1}
\end{equation}
for large $R$ and $T$, the inter-monopole potential becomes constant $2 \alpha$
in the infinite limit of $T$,
\begin{equation}
V_{M}(R) \to \lim_{T \to \infty} \frac{2 \alpha}{T}R + 2\alpha = 2 \alpha.
\label{eq:per-v1}
\end{equation}
However, in the actual lattice calculation,
we have to take a finite length of $T$, and 
hence,
the linear part $(2\alpha /T)R$ remains as a 
lattice artifact.
Therefore,
it is necessary to subtract this lattice artifact $(2\alpha /T)R$
for evaluating of the inter-monopole potential $V_M(r)$
from $W_D(R,T)$ in the lattice QCD simulation. 
The parameter $\alpha$ can be estimated from the slope of 
the dual Wilson loop $\ln \langle W_{D}(R \times T) \rangle_{\rm MA}$
for large $R$ and $T$ for each lattices.
Here, we obtain $\alpha \simeq 0.14$GeV 
from the dual Wilson loop in Fig.\ref{fig:dw} 
for $R,T > 3$ in the lattice unit.
After the subtraction of the lattice artifact $(2\alpha /T)R$,
we consider the shape of the inter-monopole potential $V_M(r)$
 in the lattice QCD
in the physical unit with $r \equiv aR$.
The inter-monopole potential $V_M(r)$ is short-ranged and flat
as shown in Fig.\ref{fig:poten}.

Here, we compare the inter-monopole potential $V_M(r)$
with the Yukawa potential
\begin{equation}
V_{Y}(r) = -\frac{(e/2)^2}{4\pi}\frac{e^{-m_B r}}{r}.
\label{eq:y-p}
\end{equation}
Since the Yukawa potential $V_Y(r)$ satisfies the relation
$\ln\{r V_Y(r)\}=-m_B r + const$, we show the logarithm plot of $r V_M(r)$
as the function of $r$ in Fig.\ref{fig:lnrV}.
In the long-distance region,
$\ln\{r V_M(r)\}$ seems to decrease linearly with $r$.
From the linear slope of $\ln\{r V_M(r)\}$
in the region of $r ~{\large ^>\!\!\!_\sim}~ 0.4$ fm,
the effective mass of the dual gluon $B_\mu$ is estimated as
$m_B \simeq 0.5$ GeV.
As shown in Fig.\ref{fig:poten2}, the inter-monopole potential
can be fitted  by the Yukawa potential $V_Y(r)$
in the long distance region using the dual gluon mass $m_B = 0.5{\rm GeV}$ and
the gauge coupling $e=2.5$.

Finally, we consider the possibility of 
the monopole size effect,
because the monopole is 
expected to be
a soliton like object composed
of gluons.
In fact, from the recent
lattice QCD study,
QCD monopoles include large off-diagonal gluons
component in their internal region even in the MA gauge 
\cite{YKIS_Suganuma,HI_HS,INNOCOM_Ichie}.
We introduce the effective size $R_{^{_{\rm M}}}
$ of the QCD-monopole, and
assume the Gaussian-type distribution of the magnetic charge
around its center,
\begin{equation}
\rho({\bf x};R_{^{_{\rm M}}}) 
= \frac{1}{(\sqrt{\pi}R_{^{_{\rm M}}})^3}\exp\left({\frac{-\vert{\bf 
x}\vert^2}{R_{^{_{\rm M}}}^2}}\right).
\label{eq:mc-den}
\end{equation}
Since the monopole part is an abelian system,
simple superposition principle on $B_\mu$
is applicable like the Maxwell equation.
Therefore, the Yukawa-type potential $V_M(r)$ 
with the effective size $R_{^{_{\rm M}}}$
of the monopole   
is expected to be
\begin{equation}
V({\bf x};R_{^{_{\rm M}}}) = 
-\frac{(e/2)^2}{4\pi}\int d^3x_1 \int d^3x_2 \rho({\bf x}_1;R_{^{_{\rm M}}})
\rho({\bf x}_2;R_{^{_{\rm M}}})
\frac{\exp({-m_B\vert{\bf x-x_1+x_2}\vert})}{
\vert{\bf x-x_1+x_2}\vert},
\label{eq:y-t-p}
\end{equation}
\noindent
or equivalently
\begin{eqnarray}
V(r;R_{^{_{\rm M}}}) = -\frac{(e/2)^2}
{\pi^2 R_{^{_{\rm M}}}^6} \int^\infty_0 dr_1 \int^\infty_0
 dr_2 e^{-(r_1^2+r_2^2)/R_{^{_{\rm M}}}^2}
\int^\pi_0 d\theta_1 \int^\pi_0 d\theta_2
\sin\theta_1\sin\theta_2 \nonumber \\
\times
\frac{\exp\left[-m_B\sqrt{ \{ \sqrt{(r-r_2\cos\theta_2)^2+(r_2\sin\theta_2)^2}
-r_1\cos\theta_1 \}^2+(r_1\sin\theta_1)^2}\right]} 
{\sqrt{ \{ \sqrt{(r-r_2\cos\theta_2)^2+(r_2\sin\theta_2)^2}
-r_1\cos\theta_1 \}^2+(r_1\sin\theta_1)^2}} 
\label{eq:y-t-p2}
\end{eqnarray}
where $r \equiv |{\bf x}-{\bf y}|$
 is the distance between the two monopole centers.
We apply this potential $V(r;R_{^{_{\rm M}}})$
to the inter-monopole potential $V_M(r)$ 
in Fig.\ref{fig:poten2}. 
The Yukawa-type potential $V(r;R_{^{_{\rm M}}})$ with the effective
monopole size $R_{^{_{\rm M}}}=0.21$fm seems to fit
the inter-monopole potential well in the whole region of the 
distance $r$.

Thus, we estimate the dual gluon mass $m_B \simeq 0.5$GeV
and the effective monopole size $R_{^{_{\rm M}}} \simeq 0.2$fm
by evaluating the inter-monopole potential $V_M(r)$
from $W_D(R,T)$ in the monopole part in the MA gauge.
We find the effective mass acquirement of the dual gluon,
which is essential for the dual Higgs mechanism
in the dual superconductor scenario,
and suggest that color confinement  
is caused by the monopole condensation
in the QCD vacuum.
The monopole size $R_{^{_{\rm M}}}$ would 
provide the critical scale for the nonperturbative QCD
in terms of the dual Higgs theory,
because off-diagonal gluons contribute to the physics
and AP-QCD should be modified at the shorter
scale than $R_{^{_{\rm M}}}$,
as well as the structure of 
the 't~Hooft-Polyakov monopole \cite{Cheng}.


\newpage
\section{Dual Gluon Propagator and Dual Gluon Effective Mass}

In the previous section, we estimate the dual gluon mass $m_B$
using the inter-monopole potential.
The more detailed measurement of $m_B$
from the inter-monopole potential
seems difficult because it requires the twice logarithmic-part
procedure from the dual Wilson loop 
\cite{INNOCOM_Tanaka,YKIS_Suganuma}.
As alternative method,
we investigate the dual gluon mass $m_B$
through the dual gluon propagator
in the lattice QCD.

\subsection{Massive Gauge Field and its Propagator}
In this subsection,
we consider the general argument on
the analytical relation between
the massive gauge field and its propagator.
To this end, we idealize here the dual gluon field $B_\mu$ as
a massive vector boson.

The lagrangian 
for the free
massive vector field $B_\mu$ in the Proca formalism 
is written as
\begin{eqnarray}
{\cal L}_{\rm Proca} 
=-\frac{1}{4}(\partial_\mu B_\nu - \partial_\nu B_\mu)^2
-\frac{1}{2}m_B^2B_\mu^2
\label{eq:LProca}
\end{eqnarray}
in the Euclidean metric \cite{Huang}.
The Proca lagrangian Eq.(\ref{eq:LProca})
leads to 
the field equation 
\cite{Itzykson}
\begin{equation}
\partial_\mu (\partial_\mu B_\nu - \partial_\nu B_\mu)-m_B^2 B_\nu=0.
\label{eq:Proca}
\end{equation}
Taking the divergence of this equation, we find 
\begin{eqnarray}
	m_B^2\partial_\mu B_\mu=0
\end{eqnarray}
For the massive case $m_B \ne 0$, $B_\mu$ is divergenceless 
as $\partial_\mu B_\mu=0$
and Eq.(\ref{eq:Proca})
reduces to the Klein-Gordon equation
\begin{equation}
(\Box - m_B^2 ) B_\mu=0. 
\label{eq:massive}
\end{equation}
The vanishing of $\partial_\mu B_\mu$ means that one of the four 
degrees of freedom of $B_\mu$ is eliminated in a covariant way.
In this case,
the propagator $\tilde{G}^D_{\mu\nu}(k)$ 
of the massive vector boson $B_\mu$ is derived as
\begin{eqnarray}
\tilde{G}^D_{\mu\nu}(k)
&\equiv& \frac{1}{k^2+m_B^2}
 	\left( \delta_{\mu\nu}+\frac{k_\mu k_\nu}{m_B^2} \right)
\end{eqnarray}
in the momentum representation \cite{Itzykson}.
The propagator ${G}^D_{\mu\nu}(x)$ in the coordinate space is 
obtained by performing the Fourier transformation as
\begin{eqnarray}
{G}^D_{\mu\nu}(x-y) \equiv \langle B_\mu(x) B_\nu(y) \rangle
	&=&
	\int \frac{d^4 k}{(2\pi)^4} e^{i k \cdot (x-y)} 
		\tilde{G}^D_{\mu\nu}(k) \nonumber \\
	&=&
	\int \frac{d^4 k}{(2\pi)^4} e^{i k \cdot (x-y)} 
	 \frac{1}{k^2+m_B^2}
 	\left( \delta_{\mu\nu}+\frac{k_\mu k_\nu}{m_B^2}\right).
\end{eqnarray}
Here, we consider the scalar-type propagator,
\begin{eqnarray}
{G}^D_{\mu\mu}(x-y) &=& \langle B_\mu(x) B_\mu(y) \rangle
	=\int\frac{d^4 k}{(2\pi)^4} e^{i k \cdot (x-y)}
	 \frac{1}{k^2+m_B^2}
 	\left( 4+\frac{k^2}{m_B^2}\right) \nonumber\\
	&=&\int\frac{d^4k}{(2\pi)^4} e^{i k \cdot (x-y)}
	\left( \frac{3}{k^2+m_B^2} + \frac{1}{ m_B^2} \right)\nonumber\\   
	&=&
	3\int\frac{d^4k}{(2\pi)^4} e^{i k \cdot (x-y)}\frac{1}{k^2+m_B^2}    
	 + \frac{1}{m_B^2}{\delta^4(x-y)} ~\label{eq:scalar-p},
\end{eqnarray}
because ${G}^D_{\mu\mu}(x-y)$ depends only 
on the four-dimensional distance $r_{\!\!_E} \equiv |x-y|$ and
it is convenient to examine $G^D_{\mu\mu}(r_{\!\!_E})$ for 
estimating the mass of the field $B_\mu$.
The integration in Eq.(\ref{eq:scalar-p}) is found to be
expressed with the modified Bessel function $K_1(z)$,
\begin{eqnarray}
I(r_{\!\!_E})	&\equiv& 
	\int\frac{d^4k}{(2\pi)^4} e^{i k \cdot (x-y)}
	\frac{1}{k^2+m_B^2}  
	=
	\frac{1}{(2\pi)^4}\int_0^\infty 
	dk \int_0^\pi d\theta \int_0^\pi d\phi \int_0^{2\pi} d\xi 
	\left( \frac{k^3 \sin^2\theta \sin\phi}
	{k^2+m_B^2}e^{ik r_{\!\!_E} \cos\theta} \right) \nonumber\\
	&=&
	\frac{2}{(2\pi)^3}\int_0^\infty dk
	\frac{k^3 \sin^2\theta}{k^2+m_B^2} 
	\left( \int_0^\pi d\theta  
	 \sin^2\theta e^{ik r_{\!\!_E} \cos\theta} \right)\nonumber\\ 
	&=&
	\frac{2}{(2\pi)^3}\int_0^\infty dk
	\frac{k^3}{k^2+m_B^2} 
	\left( \int_{-1}^{1} dt
	 \sqrt{1-t^2} e^{ik r_{\!\!_E} t} \right) 
	 =
	\frac{1}{(2\pi)^2}\int_0^\infty dk
	\frac{k^2 }{k^2+m_B^2} 
	J_1(k r_{\!\!_E})	\nonumber\\
	&=&
	\frac{1}{(2\pi)^2}\frac{m_B}{r_{\!\!_E}} K_1(m_B r_{\!\!_E}).
\end{eqnarray}
Here, we use the integration formula for the Bessel function $J_1(z)$,
\begin{eqnarray}
J_1(z) = 
	 \int_{-1}^{1} dt
	 \sqrt{1-t^2} e^{iz t},
\end{eqnarray}
and the formula for the modified Bessel function,
\begin{eqnarray}
z K_1(z) = \int_0^\infty dt
	\frac{t^2 }{t^2+z^2} 
	J_1(t).
\end{eqnarray}
Thus, the scalar-type propagator $G^D_{\mu\mu}(r_{\!\!_E})$ is written as
\begin{eqnarray}
G^D_{\mu\mu}(r_{\!\!_E})
	&=&
	\frac{3}{4\pi^2}\frac{m_B}{r_{\!\!_E}}K_1(m_B r_{\!\!_E})
+\frac{1}{m_B^2}\delta^4(x).
					\label{eq:scalar-p2}
\end{eqnarray}
Using the asymptotic expansion 
\begin{eqnarray}
K_1(m_B r_{\!\!_E}) \simeq
	\sqrt{\frac{\pi}{2m_B r_{\!\!_E}}} e^{-m_B r_{\!\!_E}} 
	\sum^\infty_{n=0}
\frac{\Gamma(\frac{3}{2}+n)}{n!\Gamma(\frac{3}{2}-n)} \frac{1}{(2m_B r_{\!\!_E})^n},
\end{eqnarray}
the scalar-type propagator Eq.(\ref{eq:scalar-p2}) reduces to
\begin{eqnarray}
G^D_{\mu\mu}(r_{\!\!_E})
&\simeq&
\frac{3\sqrt{m_B}}{2(2\pi)^{\frac{3}{2}}} \frac{e^{-m_B r_{\!\!_E}}}{r_{\!\!_E}^\frac{3}{2}}
						\label{eq:scalar-p3}
\end{eqnarray}
at the long-range as $r_{\!\!_E} > m_B^{-1}$,
since the modified Bessel function satisfies
the relation $\frac{1}{z}K_1(z) \simeq \sqrt{\frac{\pi}{2}}z^{-3/2}e^{-z}$
for $z>1$
as shown in Fig.\ref{fig:Bessel2}.
In the Eq.(\ref{eq:scalar-p3}), the damping factor $e^{-m_B r_{\!\!_E}}$ 
expresses the short-range interaction
mediated by the massive field in the coordinate space.
Then,
the mass $m_B$ of the vector field $B_\mu(x)$ is estimated 
from the slope in the logarithmic plot of 
$\frac{r_{\!\!_E}^\frac{3}{2}}{\sqrt m_B }G^D_{\mu\mu}(r_{\!\!_E})$ as
the function of $r_{\!\!_E}$,
\begin{eqnarray}
\ln \left\{ {\frac{r_{\!\!_E}^\frac{3}{2}}{\sqrt m_B }G^D_{\mu\mu}(r_{\!\!_E})} \right\} 
	&\simeq& 
	{-m_B r_{\!\!_E}}+const.
					\label{eq:ln-propa}
\end{eqnarray}


\subsection{Numerical Results from Lattice QCD Study}

In this subsection, we show the numerical results of 
the lattice QCD simulation on the propagator and 
the effective mass of the dual gluon $B_\mu$.
For the study of the dual Higgs mechanism in QCD,
we numerically 
calculate the dual gluon propagator 
$G^D_{\mu\mu}(r_{\!\!_E}) \equiv \langle B_\mu(x) B_\mu(x) \rangle_{\rm MA}$
and try to estimate of the effective mass $m_B$ 
of the dual gluon field $B_\mu$ 
in the MA gauge
using the SU(2) lattice with $24^4$ and $\beta=2.4,2.45$.
All measurements are performed at every 200 sweeps
after a thermalization of 10000 sweeps using 
the heat-bath algorithm.
We prepare 60 samples of gauge configurations.
These simulations have been performed using NEC SX-4 at 
Computation Center of Osaka University.

For the effective-mass acquirement of the dual gluon
$B_\mu$ as the result of infrared monopole condensation,  
we evaluate the dual gluon propagator
obtained from the lattice QCD in the MA gauge.
Since 
the massive vector-boson propagator is expected to
behave as
$G^D_{\mu\mu}(r_{\!\!_E})\sim r_{\!\!_E}^{-3/2}\exp (-m_B r_{\!\!_E})$ 
at the long distance
as shown in Eq.(\ref{eq:ln-propa}),
we investigate the logarithm plot of ${r_{\!\!_E}}^{3/2}G_{\mu\mu}^{D}(r_{\!\!_E})$
as the function of $r_{\!\!_E}$
for the estimation 
of the dual gluon mass $m_B$.

We show in Figs.\ref{fig:lnGB},\ref{fig:GBfit}
the lattice QCD data for the dual gluon correlations,
$\ln \{r_{\!\!_E}^{3/2} G^D_{\mu\mu}(r_{\!\!_E}) \}$
and $G^D_{\mu\mu}(r_{\!\!_E})$,
as the function of the 4-dimensional Euclidean
distance $r_{\!\!_E}$.
In Fig.\ref{fig:lnGB},
the dual gluon correlation
$\ln \{ r_{\!\!_E}^{3/2} G^D_{\mu\mu}(r_{\!\!_E}) \}$
decreases linearly with $r_{\!\!_E}$ in the region of 
$r_{\!\!_E} ~{\large^>\!\!\!_\sim}~ 0.8$ fm.
We try to fit a straight line to the lattice QCD data 
of $\ln \{ r_{\!\!_E}^{3/2} G^D_{\mu\mu}(r_{\!\!_E}) \}$
in this region,
and estimate the effective mass $m_B$ of the dual gluon $B_\mu$
as $m_B \simeq 0.4$ GeV
($m_B \simeq 0.41$ GeV  for $\beta=2.4$
and $m_B \simeq 0.38$ GeV for $\beta=2.45$).
In Fig.\ref{fig:lnGB}(b),
based on Eq.(\ref{eq:ln-propa}),
we compare 
the dual gluon correlation
$\ln \left( r_{\!\!_E}^{3/2} G^D_{\mu\mu}(r_{\!\!_E}) \right)$
with the massive vector-boson correlation
with $m_B=0.4$ GeV in the long distance as 
$r_{\!\!_E} ~{\large^>\!\!\!_\sim}~ 0.8$ fm.
We find good agreement between the lattice QCD data
and the massive boson case denoted by the dotted line,
and then the dual gluon field $B_\mu$ is consider to have
the effective mass $m_B \simeq 0.4$ GeV in the infrared region.

In Fig.\ref{fig:GBfit},
we show the scalar-type dual gluon propagator
$G^D_{\mu\mu}(r_{\!\!_E}) \equiv 
\langle B_\mu(x) B_\mu(y) \rangle_{\rm MA}$
in the linear scale.
The dotted curve denotes the analytical form
in Eq.(\ref{eq:scalar-p2})
for the massive vector-boson propagator with $m_B=0.4$ GeV.
In the long-distance region as $r_{\!\!_E}~{\large ^>\!\!\!_\sim} 0.8$
fm, we find
\begin{eqnarray}
G^D_{\mu\mu}(r_{\!\!_E})
        &\simeq&
        const \cdot \frac{m_B}{r_{\!\!_E}}K_1(m_B r_{\!\!_E}),
\end{eqnarray}
and the dual gluon $B_\mu$ seems to propagate
as the massive vector field with $m_B \simeq 0.4$ GeV.
On the other hand,
in the short-distance region,
$G^D_{\mu\mu}(r_{\!\!_E})$
seems to differ from the massive vector propagator.
We speculate that
such a short-distance deviation on $G^D_{\mu\mu}(r_{\!\!_E})$
is brought by the monopole-size effect,
because the monopole is considered as a soliton-like object
composed of gluons \cite{INNOCOM_Tanaka}.

In this way, we find the effective mass $m_B \simeq 0.4$ GeV
of the dual gluon $B_\mu$ from the infrared behavior of 
the dual gluon propagator $G^D_{\mu\nu}\equiv \langle B_\mu(x) B_\nu(y)
\rangle_{\rm MA}$ in the MA gauge using the lattice QCD simulation.
The effective-mass acquirement of the dual gluon $B_\mu$
supports
the dual Higgs mechanism by monopole condensation in the QCD vacuum. 

\newpage
\section{Summary and Concluding Remarks}
In order to clarify the dual superconductor picture
for the quark confinement mechanism  
in the QCD vacuum, we 
have studied
the effective-mass acquirement of the dual gluon field,
which is essential for the dual Higgs mechanism.
In the maximally abelian (MA) gauge,
QCD is reduced into an abelian gauge theory
including both the color-electric current and the 
color-magnetic monopole current.

For the investigation of the dual Higgs mechanism 
by monopole condensation in QCD, 
we have introduced the dual gluon field $B_\mu$
and have studied its features 
in the monopole part (the monopole-current system)
using the MA gauge in the lattice QCD Monte Carlo simulation.
Owing to the absence of the electric current,
the monopole part resembles the dual version of QED,
and hence this system
is naturally described by the dual gluon field $B_\mu$
without meeting the difficulty on the Dirac-string singularity.
In the dual gauge formalism, the dual Higgs mechanism
is characterized by the acquirement of the effective mass $m_B$
of the dual gluon field $B_\mu$.
Then, we have investigated the dual gluon mass $m_B$
by examining the inter-monopole potential and the dual gluon propagator,
respectively.

As the first attempt
to evaluate the dual gluon mass,
we have calculated the dual Wilson loop
$\langle W_D(R \times T) \rangle_{\rm MA}$, and have studied the 
inter-monopole potential $V_M(r)$
in the monopole part in the MA gauge
by using the lattice QCD simulation.
In the lattice QCD,
we have found that 
the dual Wilson loop
obeys the perimeter law for large loops.
Considering the finite-size effect of the dual Wilson loop,
we have numerically derived the inter-monopole potential $V_M(r)$,
and have found that
$V_M(r)$ is short-ranged and flat
in comparison with the linear inter-quark potential.
Then, we have compared 
the inter-monopole potential $V_M(r)$ with the Yukawa potential and 
have estimated the dual gluon mass as
$m_B \simeq {\rm 0.5GeV}$, 
which is consistent with 
the phenomenological parameter fitting 
in the dual Ginzburg-Landau theory \cite{SST}.
The generation of the dual gluon mass $m_B$ supports 
the realization of the dual Higgs mechanism 
and monopole condensation
in the QCD vacuum.
To explain the short-range deviation between 
the inter-monopole potential $V_M(r)$
and the Yukawa potential,
we have considered the effective size $R_{^{_{\rm M}}}$ of the monopole,
which is considered as a soliton-like object composed of gluons.
We have found a good agreement of $V_M(r)$ and
the Yukawa-type potential $V(r;R_{^{_{\rm M}}})$
with the monopole effective size $R_{^{_{\rm M}}}$ in the whole region of $r$.
The monopole size has been estimated as $R_{^{_{\rm M}}} \simeq {\rm 0.2fm}$, which would 
provide the critical scale for the NP-QCD
in terms of the dual Higgs theory, because
the monopole should be treated as a non-local soliton
at the shorter scale than $R_{^{_{\rm M}}}$.

As the second attempt,
we have evaluated the 
effective dual-gluon mass
by investigating the dual gluon propagator
$G^D_{\mu\mu}(x-y) \equiv \langle B_\mu(x) B_\mu(x) \rangle_{\rm MA}$
in the MA gauge in the lattice QCD.
The dual gluon propagator
seems to be well fitted by
the massive vector-boson propagator
at the long distance.
From the behavior of the dual gluon propagator
in the infrared region,
we have estimated the effective dual-gluon mass $m_B \simeq 0.4$ GeV.

Thus, we have shown 
the effective mass acquirement of the dual gluon field $B_\mu$
in the studies of the inter-monopole potential and the dual gluon propagator
in the MA gauge using the lattice QCD simulation.
This result support the dual Higgs mechanism
induced by the monopole condensation, 
which would be responsible for quark confinement,
at the infrared scale of QCD.
The values of the dual gluon mass obtained by the two methods
seem to coincide within the numerical error.
Here, $m_B^{-1} \sim 0.4$ fm corresponds to the ``penetration depth'' 
of the dual superconductor
and provides the radius of the color-electric flux tube of hadrons 
\cite{SST}.
Then, the estimation of the dual gluon mass,
$m_B = 0.4 \sim 0.5$ GeV,
determines the most relevant quantity of the dual superconducting
theory
or the key parameter of 
the dual Ginzburg Landau (DGL) theory \cite{SST}. 

\section*{Acknowledgment}
We would like to thank Professor Hiroshi Toki
for his useful comments and discussions.
One of authors (H.S.) is supported in part by Grant for
Scientific Research (No.09640359) from the Ministry of Education,
Science and Culture, Japan.
The lattice QCD simulations have been performed on the super-computer
SX4 at Osaka University.


\baselineskip 1cm

\begin{figure}
          \caption{\baselineskip .9cm
The dual Wilson loop $\langle W_D(R,T) \rangle_{\rm MA}$
as the function of 
its perimeter $L \equiv 2(R+T)$ in the monopole part
in the MA gauge in the SU(2) lattice QCD
with $20^4$ lattice and $\beta=2.3$.
The data for the large loops with $R,T>3$
are shown.
The perimeter law seems to hold for
$\langle W_D(R,T) \rangle_{\rm MA}$.
\label{fig:dw}}      
\end{figure}

\begin{figure}
\baselineskip 1cm
          \caption{\baselineskip .9cm
The inter-monopole potential
$V_M(r)$ as the function of the distance $r$ between the monopoles
in the monopole part in the MA gauge in the SU(2) lattice QCD
with $20^4$ lattice and $\beta=2.207, 2.257, 2.3$.
	\label{fig:poten}}

\end{figure}

\baselineskip 1cm
\begin{figure}
\caption{\baselineskip .9cm
The logarithmic plot for $r V_M(r)$
as the function of $r$
in the monopole part in the MA gauge in the SU(2) lattice QCD
with $20^4$ and $\beta=2.207, 2.257, 2.3$.
In the long-distance region,
$\ln\{r V_M(r)\}$
seems to decrease linearly with $r$.
For comparison,
We plot also the solid line $\ln\{r V_M(r)\}
=-m_B r + const$ with $m_B=0.5$ GeV.
The effective mass of the dual gluon $B_\mu$ is estimated as
$m_B \simeq 0.5$ GeV from the linear slope of
$\ln\{r V_M(r)\}$
in the region of $r ~{\large ^>\!\!\!_\sim}~ 0.4$ fm.
\label{fig:lnrV}
}
\end{figure}

\begin{figure}
\baselineskip 1cm
\caption{\baselineskip .9cm
The inter-monopole potential $V_M(r)$ as the function 
of the distance $r \equiv |{\bf x}-{\bf y}|$ between the monopoles
in the SU(2) lattice QCD with $20^4$ and 
$\beta=2.207, 2.257, 2.3$.
The dotted and the solid curves denote
the simple Yukawa potential $V_{\rm Y}(r)$ 
and the Yukawa-type potential $V(r;R_{^{_{\rm M}}})$ with the 
monopole size effect, respectively. 
The Yukawa-type potential $V(r;R_{^{_{\rm M}}})$ seems to
fit the inter-monopole potential $V_M(r)$
with the effective monopole size $R_{^{_{\rm M}}} \simeq {\rm 0.2fm}$
in the whole region of r. 
\label{fig:poten2}}
\end{figure}

\baselineskip 1cm
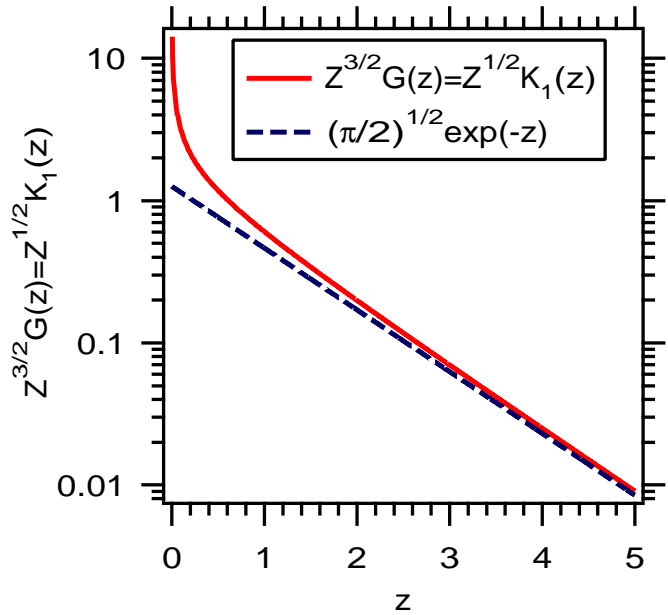
\begin{figure}
          \caption{\baselineskip .9cm
(a) The function $G(z) \equiv \frac{1}{z}K_1(z)$
(solid curve) appearing in the massive boson propagator and
its asymptotic form $\sqrt{\frac{\pi}{2z^3}}e^{-z}$ (dotted curve)
as the function of $z$.
(b) The logarithmic plots for $z^{3/2}G(z)=\sqrt{z}K_1(z)$ (solid curve) and 
$\sqrt{\frac{\pi}{2}}e^{-z}$ (dotted line)
as the function of $z$. 
For $z > 1$, the function $G(z)$ is apporoximated 
by its asymptotic form as 
$G(z) \simeq \sqrt{\frac{\pi}{2z^3}}e^{-z}$.
	\label{fig:Bessel2}}

\end{figure}
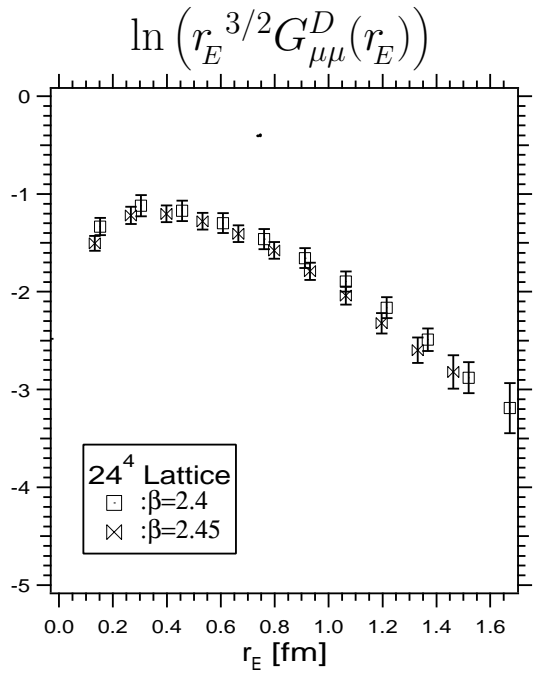
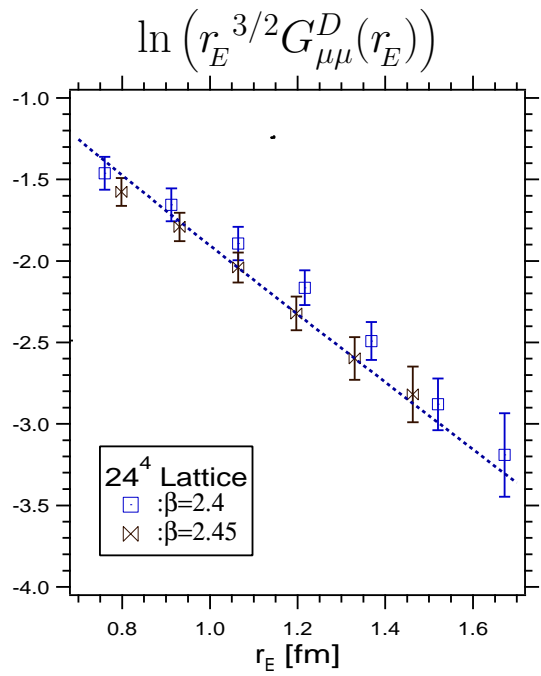
\begin{figure}

\baselineskip 1cm
          \caption{\baselineskip .9cm
(a) The logarithmic plot of ${r_{\!\!_E}}^{3/2}G_{\mu\mu}^{D}(r_{\!\!_E})$
as the function of $r_{\!\!_E}$
in the monopole part in the MA gauge in the SU(2) lattice QCD
with $24^4$ and $\beta=2.4, 2.45$.
The correlation
$\ln \left( r_{\!\!_E}^{3/2} G^D_{\mu\mu}(r_{\!\!_E}) \right)$
decreases linearly with $r_{\!\!_E}$
in the long-distance region.
(b) The comparison with the analytical form in Eq.(\ref{eq:scalar-p2})
for the massive vector-boson propagator with the mass $m_B =0.4$ GeV.
The massive-boson correlation denoted by the dotted line
almost reproduces the lattice QCD data for the dual gluon correlation
$\ln \left( r_{\!\!_E}^{3/2} G^D_{\mu\mu}(r_{\!\!_E}) \right)$
at the long distance as $r ~{\large ^>\!\!\!_\sim}~ 0.8$ fm. 
The effective mass of the dual gluon is estimated as
$m_B \simeq 0.4$ GeV from the linear slope of 
$\ln \left( r_{\!\!_E}^{3/2} G^D_{\mu\mu}(r_{\!\!_E}) \right)$
in the infrared region.
	\label{fig:lnGB}}
\end{figure}
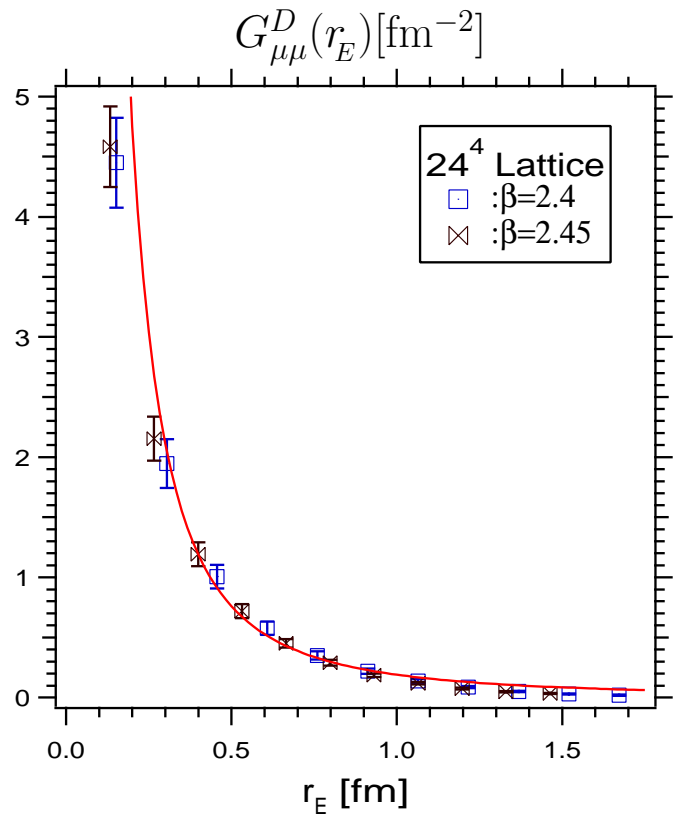
\begin{figure}

\baselineskip 1cm
          \caption{\baselineskip .9cm
The scalar-type dual gluon propagator $G^D_{\mu\mu}(r_{\!\!_E})$
as the function of the  4-dimensional Euclidean
distance $r_{\!\!_E}$
in the monopole part in the MA gauge in the SU(2) lattice QCD
with $24^4$ and $\beta=2.4, 2.45$.
The dotted curve denotes the propagator of the massive vector boson
with the mass $m_B=0.4$ GeV.
The dual gluon $B_\mu$ seems to propagate as the massive vector field
at the long distance as $r_{\!\!_E} ~{\large ^>\!\!\!_\sim}~ 0.8$ fm.
	\label{fig:GBfit}}
\end{figure}
\clearpage
\newpage
\begin{figure}[hb]
	\centerline{\Large $\langle W_D(R,T) \rangle_{\rm MA}$}
	\epsfxsize = 10cm
          \centerline{\epsfbox{dw-p.EPSF}}
	\centerline{\Huge Figure \ref{fig:dw}}
     \end{figure}
\newpage
\begin{figure}[hb]
	\centerline{\Large Inter-Monopole Potential $V_M(r)$}
          \epsfxsize = 10cm
          \centerline{\epsfbox{imp.EPSF}}
	\centerline{\Huge Figure \ref{fig:poten}}
     \end{figure}
\newpage
\begin{figure}
	\centerline{\Large $\ln \{r V_M(r)\}$}
\begin{center}
\epsfig{figure=lnrVl.EPSF,height=8cm}
\\
\vspace{1.0cm}
{\Huge Figure \ref{fig:lnrV}}
\end{center}
\end{figure}
\newpage
\begin{figure}[hb]

	\centerline{\Large Inter-Monopole Potential $V_M(r)$}
\epsfysize=8cm
\centerline{\epsfbox{imp-r02.EPSF}}
	\centerline{\Huge Figure \ref{fig:poten2}}
\end{figure}
\clearpage
\newpage
%
\begin{figure}[h]
          \centerline{\Large $G(z) \equiv \frac{1}{z}K_1(z)$}
          \epsfxsize = 9.5cm
          \centerline{\epsfbox{Bessel2.EPSF}}
\centerline{(a)}
          \centerline{\Large $z^\frac{3}{2}G(z)=\sqrt{z}K_1(z)$}
          \epsfxsize = 9.5cm
          \centerline{\epsfbox{lnbessel2.EPSF}}
\centerline{(b)}
	\centerline{\Huge Figure \ref{fig:Bessel2}}
     \end{figure}
\newpage
\begin{figure}[htb]
	\centerline{~~~\Large$\ln \left( {r_{\!\!_E}}^{3/2}
G^D_{\mu\mu}(r_{\!\!_E}) \right)$} 
          \epsfxsize = 8cm
          \centerline{\epsfbox{lnGB.EPSF}}
          \centerline{(a)}
	\centerline{~~~\Large$\ln \left( {r_{\!\!_E}}^{3/2}
G^D_{\mu\mu}(r_{\!\!_E}) \right)$} 
          \epsfxsize = 8cm
          \centerline{\epsfbox{lnGBfit.EPSF}}
          \centerline{(b)}
	\centerline{\Huge Figure \ref{fig:lnGB}}
     \end{figure}
\newpage
\begin{figure}[htb]
     \end{figure}
\newpage
\begin{figure}[htb]
	\centerline{~~~ \Large $ G^D_{\mu\mu}(r_{\!\!_E})
	 {\rm [fm^{-2}]}$} 
          \epsfxsize = 10cm
          \centerline{\epsfbox{GBfit.EPSF}}
	\centerline{\Huge Figure \ref{fig:GBfit}}
     \end{figure}
\clearpage
\newpage
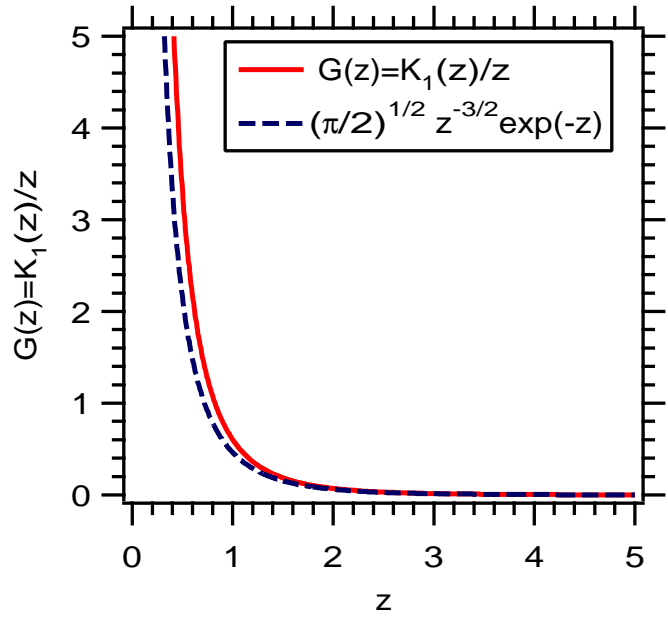
\end{document}